\documentclass[aps,pra,preprintnumbers,twoside,twocolumn,superscriptaddress,floatfix,10pt]{revtex4-1}

\usepackage[version=3]{mhchem} 
\usepackage{amsmath}
\usepackage{amssymb}
\usepackage{amsthm}
\usepackage{amsfonts}
\usepackage{amsbsy}
\usepackage{graphicx}

\begin{document} 

\author{Birger Horstmann}\email{birger.horstmann@dlr.de}
\affiliation{Massachusetts Institute of Technology, 77 Massachusetts Avenue, Cambridge, MA 02139, USA} 
\affiliation{German Aerospace Center, Pfaffenwaldring 38-40, 70569 Stuttgart, Germany} 
\affiliation{University of Stuttgart, Pfaffenwaldring 6, 70550 Stuttgart, Germany} 
\affiliation{Helmholtz Institute Ulm, Albert-Einstein-Allee 11, 89069 Ulm, Germany} 
\author{Betar Gallant}
\affiliation{Massachusetts Institute of Technology, 77 Massachusetts Avenue, Cambridge, MA 02139, USA} 
\author{Robert Mitchell}
\affiliation{Massachusetts Institute of Technology, 77 Massachusetts Avenue, Cambridge, MA 02139, USA} 
\author{Wolfgang G. Bessler}
\affiliation{Offenburg University of Applied Sciences, Badstra\ss e 24, 77652 Offenburg, Germany} 
\author{Yang Shao-Horn}
\affiliation{Massachusetts Institute of Technology, 77 Massachusetts Avenue, Cambridge, MA 02139, USA} 
\author{Martin Z. Bazant}\email{bazant@mit.edu}
\affiliation{Massachusetts Institute of Technology, 77 Massachusetts Avenue, Cambridge, MA 02139, USA} 

\title{ Rate-dependent morphology of \ce{Li2O2} growth in \ce{Li}-\ce{O2} batteries}

\begin{abstract}
Compact solid discharge products enable energy storage devices with high gravimetric and volumetric energy densities, but solid deposits on active surfaces can disturb charge transport and induce mechanical stress.  In this Letter we develop a nanoscale continuum model for the growth of \ce{Li2O2} crystals in lithium-oxygen batteries with organic electrolytes, based on a theory of electrochemical non-equilibrium  thermodynamics originally applied to Li-ion batteries.  As in the case of lithium insertion in phase-separating \ce{LiFePO4} nanoparticles, the theory predicts a transition from complex to uniform morphologies of \ce{Li2O2} with increasing current. Discrete particle growth at low discharge rates becomes suppressed at high rates, resulting in a film of electronically insulating \ce{Li2O2} that limits cell performance. We predict that the transition between these surface growth modes occurs at current densities close to the exchange current density of the cathode reaction, consistent with experimental observations. 
\end{abstract}

\maketitle

{\bf Introduction.} Crystallization on active surfaces is essential in many battery and electrodeposition processes. Crystalline reaction products offer the potential for compact and lightweight energy storage, but accommodating such deposits is challenging for electrode design. The wide range of conditions during crystallization causes a multitude of growth morphologies in electrochemical systems. In lead-acid batteries, particle sizes of deposited \ce{Pb} depend on voltage sweeping rates \cite{Yamaguchi2001}; in alkaline \ce{Zn} batteries or \ce{Zn}-\ce{O2} batteries, electrodeposited \ce{ZnO}  undergoes a transition from film-growth to dendritic-growth as a function of cycling depth \cite{Yuan2006}, influenced by electrolyte additives \cite{Lee2006}; in metal electrodeposition, dendritic growth depends sensitively on the electrolyte composition and applied current~\cite{kuhn1994,rosso2007,nishikawa2010}; in rechargeable lithium batteries, morphological changes in Li metal anodes during dissolution, plating and dendritic growth~\cite{nishikawa2010}, are a critical challenge, subject to ongoing modeling efforts \cite{ISI:000185639800015,Ely2013}. 

\begin{figure}[b]
	\centering
	\includegraphics[width=1\columnwidth]{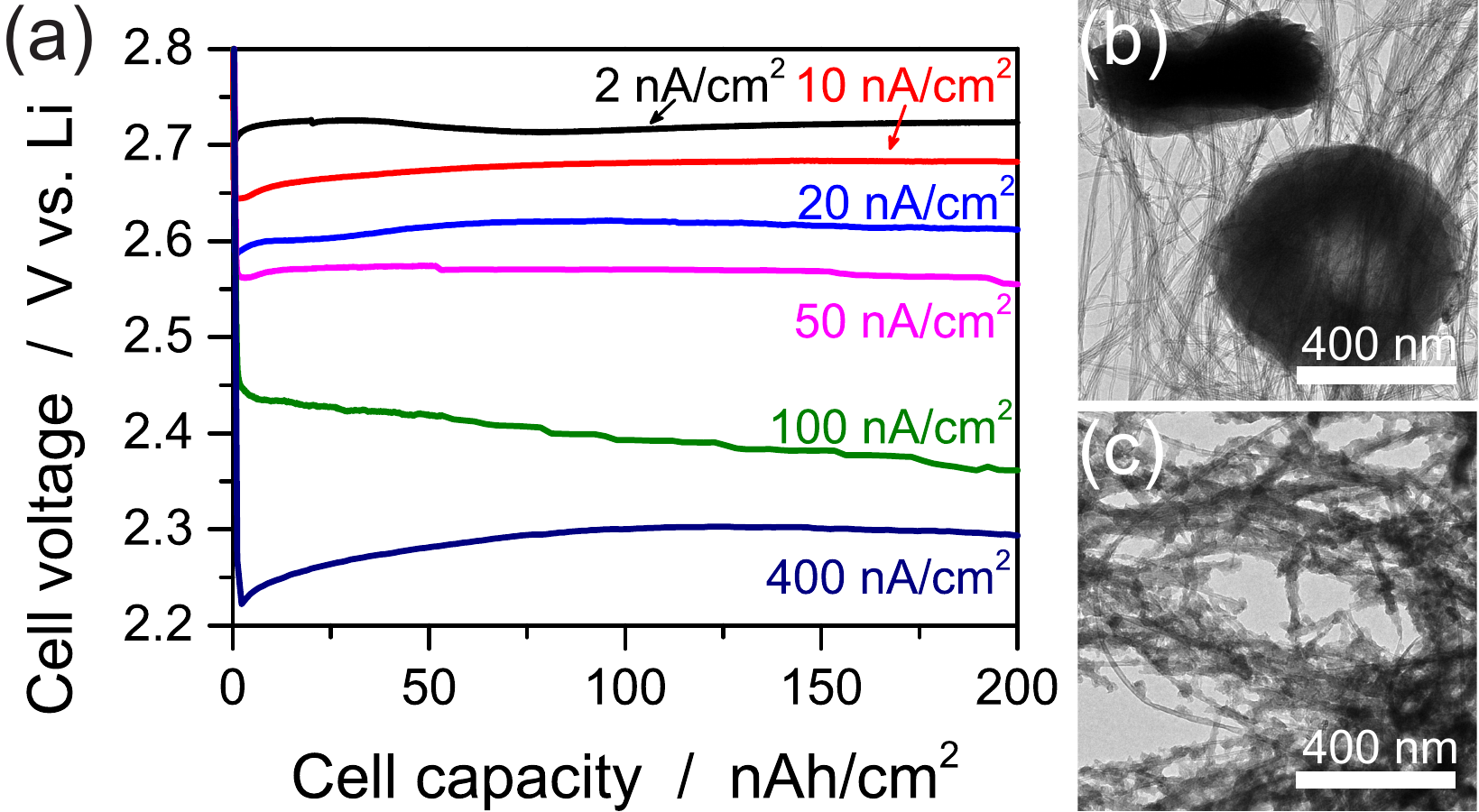}	
	\caption{Galvanostatic discharge of \ce{Li}-\ce{O2} battery with CNT cathode \cite{Gallant2012a, Mitchell2013a}. The average \ce{Li2O2} thickness at $100\text{ nAh/cm}^2$ is $1$ molecular monolayer. (a) Discharge voltage for various discharge currents. (b) TEM micrograph for $I=2 \text{ nA/cm}^2$ at $280 \text{ nAh/cm}^2$ with individual particles. (c) TEM micrograph for $I=50 \text{ nA/cm}^2$ at $840 \text{ nAh/cm}^2$ with coating by small particles. Currents are normalized to true surface area.}
	\label{fig:Exp}
\end{figure}

Recent experiments on  \ce{Li}-\ce{O2}  batteries with ether-based electrolyte have revealed that the electronically isolating discharge product \ce{Li2O2} can deposit in complex toroid-like morphologies \cite{Mitchell2011,Gallant2013} or thin films \cite{Mccloskey2012,Gallant2013}. In contrast, only quad-shaped particles have been observed in sodium-oxygen batteries so far \cite{Hartmann2012}. 
\ce{Li}-\ce{O2} batteries are prominent candidates for next-generation batteries that can replace conventional combustion technologies \cite{Abraham1996, Girishkumar2010, Christensen2012, Bruce2012a, Lu2012, Scrosati2010}. 
Although the stability of oxygen electrode and electrolyte remains a challenge for practical \ce{Li}-\ce{O2} batteries, ether-based electrolytes remain relatively stable \cite{Mccloskey2011, Lu2011a}.

The morphology of \ce{Li2O2} formed upon discharge in ether-based electrolytes has an as-yet unexplained dependence on the applied current. An evolution from single-crystalline disc to complex toroid-like morphologies during discharge was first observed in nano-structured electrodes with large surface areas \cite{Mitchell2011, Gallant2012a} (Fig. \ref{fig:Exp}).  This has since been confirmed on different carbon substrates at low surface specific rates \cite{Mitchell2011, Gallant2012a, Lu2011a, Black2012, ISI:000185639800015}. Although the disc-like particles reach 100 nm sizes, toroid-like particles can grow much larger, and the electron transport path and growth mechanisms are just beginning to be understood~\cite{Mitchell2013a}. Regardless of this complex behavior at low rates, however, \ce{Li2O2} forms a crystalline film on the active surfaces of the cathode at high surface specific rates that limits the electrode capacity and achievable power density. When the film thickness approaches $5~\text{nm}$, the active surfaces become passivated, as electronic resistance increases with thickness \cite{Viswanathan2011}. 

In this Letter we model the rate-dependent morphological transition in \ce{Li2O2} growth, using the recently developed variational theory of electrochemical kinetics~\cite{Bazant2013,Singh2008,Bai2011,Cogswell2012,Ferguson2012a,Cogswell2013a} applied to classical surface-growth models~\cite{Burton1951,Stone2005,Margetis2005}. 
The theory predicts a transition starting in the first monolayer from particle growth to film growth when the current exceeds the exchange current for the oxygen reduction reaction, consistent with experimental observations. The mechanism is analogous to the suppression of phase separation in \ce{LiFePO4} nanoparticles, first predicted by the same general theory~\cite{Bai2011,Cogswell2012,Ferguson2012a}.

{\bf Theory.} Existing models of \ce{Li}-\ce{O2} batteries are either macroscopic or atomistic. Cell-level models propose pore blocking due to reaction products \cite{Sandhu2007, Williford2009, Andrei2010, Neidhardt2012, Horstmann2012a} and surface passivation \cite{Albertus2011, Viswanathan2011}. Atomistic models discuss the surface structure of \ce{Li2O2} crystals \cite{Mo2011, Radin2012, Radin2012a,Hummelshoj2013,Viswanathan2013}, the kinetics of the oxygen reduction/evolution in aprotic electrolytes \cite{Hummelshoj2010, Mo2011,Viswanathan2013}, and the electron conductivity of \ce{Li2O2} \cite{Viswanathan2011, Ong2012, Radin2012}. Here, we develop a nanoscale continuum model based on these atomistic studies, which bridges the gap to macroscopic models by predicting morphological selection in the early stages of surface growth.

\begin{figure}[t]
	\centering
	\includegraphics[width=\columnwidth]{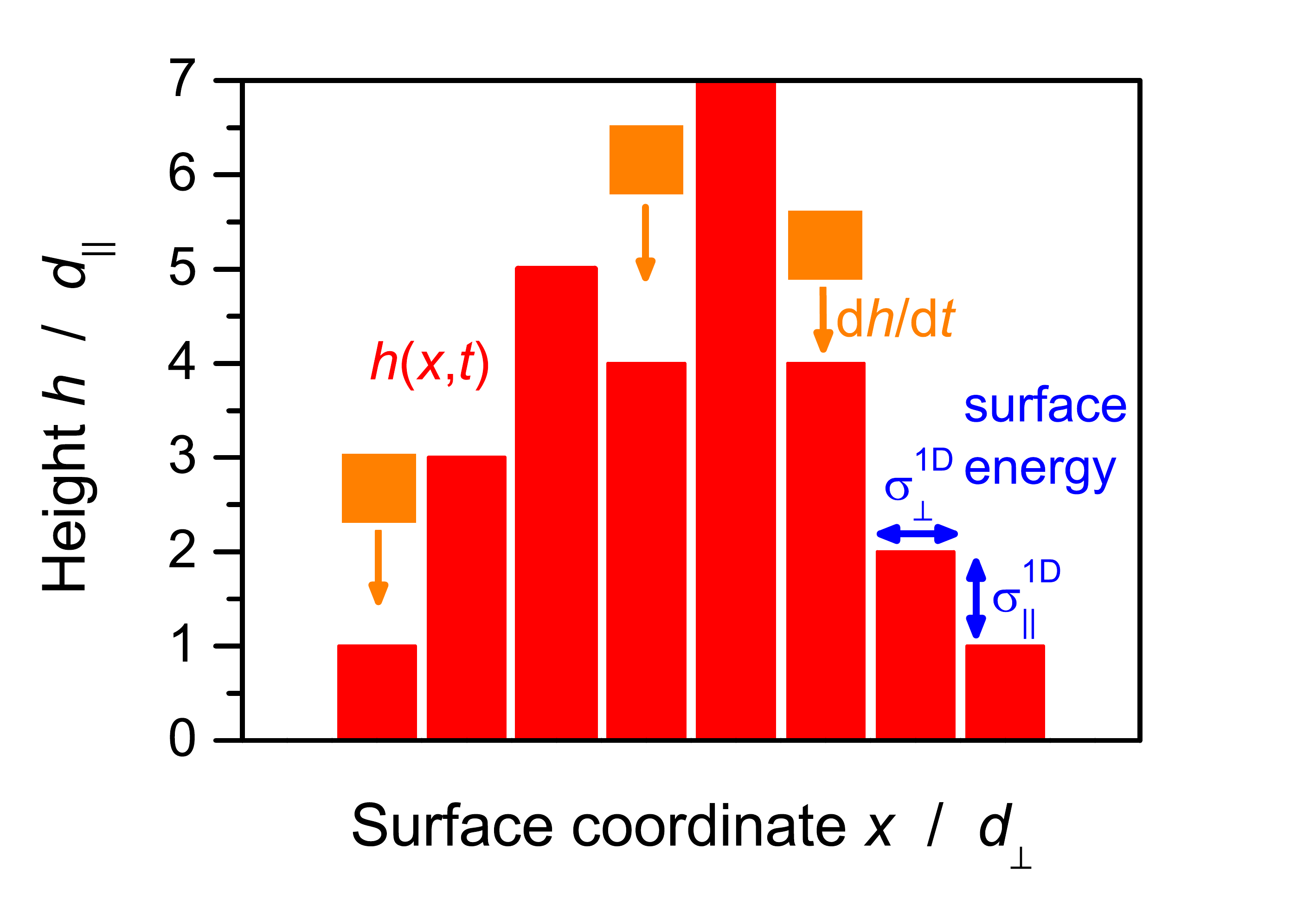}
	\caption{Scheme of the (1+1)-D surface model. Individual \ce{Li2O2} molecules are added on top of a surface crystal of height $h(x,t)$ at the rate $\partial h/\partial t$. The dimensionless variables $\tilde h =h/d_\|$ and $\tilde x=x/d_\perp$ are used for height and surface coordinate, where $d_\|$ and $d_\perp$ are the distances between molecules in the horizontal and vertical direction.}
	\label{fig:surface growth}
\end{figure}

We model the electrodeposition oxygen reduction reaction (ORR), 
\begin{equation}
\label{eq:ORR}
2\ce{Li+}+\ce{O2} + 2\ce{e-} \rightleftharpoons \ce{Li2O2}
\end{equation}
on a carbon surface in (1+1)-dimensional space, i.e., through the height of the crystal $h(x)$ as a function of the  projected surface coordinate $x$ (Fig. \ref{fig:surface growth}). In this way, \ce{Li2O2} molecules align in columns growing at the electrochemically controlled rate $\frac{\partial h}{\partial t}$. The continuous evolution of $h(x,t)$ is a standard mathematical description of surface growth~\cite{barabasi_book}.  

We choose the $\text{O-rich }(0001)$-surface for the top-facets and the $(1100)$-surface for the side-facets \cite{Hummelshoj2013}. $d_\|=0.380\text{ nm}$ and $d_\perp=0.313\text{ nm}$ are the corresponding distances between \ce{Li2O2} molecules in the bulk crystal \cite{Ong2012}. Our choice is motivated by the Wulff shape of the \ce{Li2O2} crystal, reconstructed from ab-initio simulations of the surface energies \cite{Mo2011, Radin2012, Radin2012a, Hummelshoj2013}. It agrees with microscopy of the preferred crystal orientation in disc-like and toroid-like particles \cite{Mitchell2013a}.  Integer values of $\tilde{h}=h/d_\|$ correspond to completely deposited monolayers, and non-integer values to intermediate states and partially filled layers. 

We extract the surface energies $\sigma$ from ab initio calculations \cite{Hummelshoj2013}. Our $1$D surface model is based on $\sigma_\perp^\text{1D}=\sigma_\perp A_\perp/d_\perp=140~\text{meV}/d_\perp$ and $\sigma_\|^\text{1D}=\sigma_\|6A_\|/2d_\|=540~\text{meV}/d_\|$, where $A_\perp=\sqrt{3}d_\perp^2/2$ and $A_\|=d_\perp d_\|/\sqrt{3}$ are the areas of the top-facets and side-facets of individual molecules, respectively. The predicted Wulff shape varies among different studies \cite{Mo2011, Radin2012, Radin2012a, Hummelshoj2013}, but does not affect our main result below, that the growth mode goes through a transition close to the exchange current, for any of these Wulff shapes.

We describe the current density profile $I(x,t)$ using generalized Butler-Volmer kinetics  based on non-equilibrium thermodynamics, recently developed by Bazant {\em et al.}~\cite{Bazant2013} and applied to intercalation dynamics in Li-ion batteries~\cite{Singh2008, Burch2009, Bai2011, Cogswell2012, Ferguson2012a, Cogswell2013a}.  Here, we apply the theory for the first time to surface growth, using a different model for the \ce{Li2O2} chemical potential, 
\begin{equation}
\mu = \frac{\delta G[c]}{\delta c}
=d_\|d_\perp\frac{\delta G[h]}{\delta h}   \label{eq:mu}
\end{equation}
which is the variational derivative of the Gibbs free energy $G=\int_0^L g~\text{d}x$ (defined below),
where $c(x,t)=h(x,t)/(d_\|d_\perp)$ is the concentration of \ce{Li2O2} molecules per substrate length.
We choose as reference state, where $\mu  = \mu^\Theta$, the fully charged state without any \ce{Li2O2} at room temperature, and atmospheric pressure ($h=0$, $T=298.15~\text{K}$, $p=1~\text{atm}$). The battery voltage, $E$, has the open circuit value, $E_0$ in this reference state. We assume constant activities for lithium ions, oxygen molecules, and electrons, $a_{\text{Li}^+}=a_{\text{O}_2}=a_{\text{e}^-}=1$, during morphology selection in the early stages of growth since thin \ce{Li2O2} deposits ($<15$ molecular layers) have negligible electronic resistivity~\cite{Viswanathan2011} and cause negligible electrolyte depletion at  typical currents. In equilibrium, the voltage increment, $\Delta\Phi=E-E_0$, is then given by the Nernst equation,
\begin{equation}
\Delta\Phi_\text{eq} = -\frac{k_\text{B}T}{2e}\ln a = \frac{ \mu^\Theta - \mu }{2e},   \label{eq:nernst}
\end{equation}
where $a$ is the \ce{Li2O2} activity. The variational activity, $a$, and the chemical potential, $\mu$, determine the thermodynamics of \ce{Li2O2} deposits up to a few monolayers and depend on sensitively their profile, $h(x,t)$ (see Eq. \ref{eq:mu}). 

Out of equilibrium, the two dimensional current density $I(x,t)$ (per substrate area) is given by the Butler-Volmer equation, 
\begin{equation}
\label{eq:current}
I=A\cdot I_0\left[e^{-\alpha 2e\eta/k_\text{B}T} - e^{\left(1-\alpha\right) 2e\eta/k_\text{B}T} \right],
\end{equation} 
in terms of the activation overpotential $\eta$, the exchange current density $I_0$ ~\cite{Bazant2013}, and a geometrical factor converting substrate length to normal surface length $A$ ~\cite{barabasi_book},
\begin{eqnarray}
 \eta&=&\Delta\Phi - \Delta\Phi_\text{eq}, \\
I_0 &=& \frac{2ek_0a^\alpha}{\gamma_\ddag} \label{eq:I0},\\
A &=&\sqrt{1+\left(\frac{\partial h}{\partial x}\right)^2}, \label{eq:A}
\end{eqnarray}
respectively. Note that in our model $I_0$ depends on activity, which is a complicated function of the height profile $h(x)$.
We assume that the first charge transfer step in the ORR (\ref{eq:ORR}) is rate limiting and symmetric ($\alpha_1 = \frac{1}{2}$), so the overall charge transfer coefficient is $\alpha = \frac{1}{4}$ (see also~ \cite{Mo2011, Hummelshoj2010}), which is consistent with the Tafel slope measured on glassy carbon \cite{Lu2013}. The activity coefficient of the transition state $\gamma_\ddag$ is approximately constant and can be estimated by Marcus theory~\cite{Bazant2013} because it is dominated by desolvation. Setting $\gamma_\ddag=1$, the rate constant $k_0$ is determined by Tafel analysis below.  

\begin{figure}[t]
	\centering
	\includegraphics[width=\columnwidth]{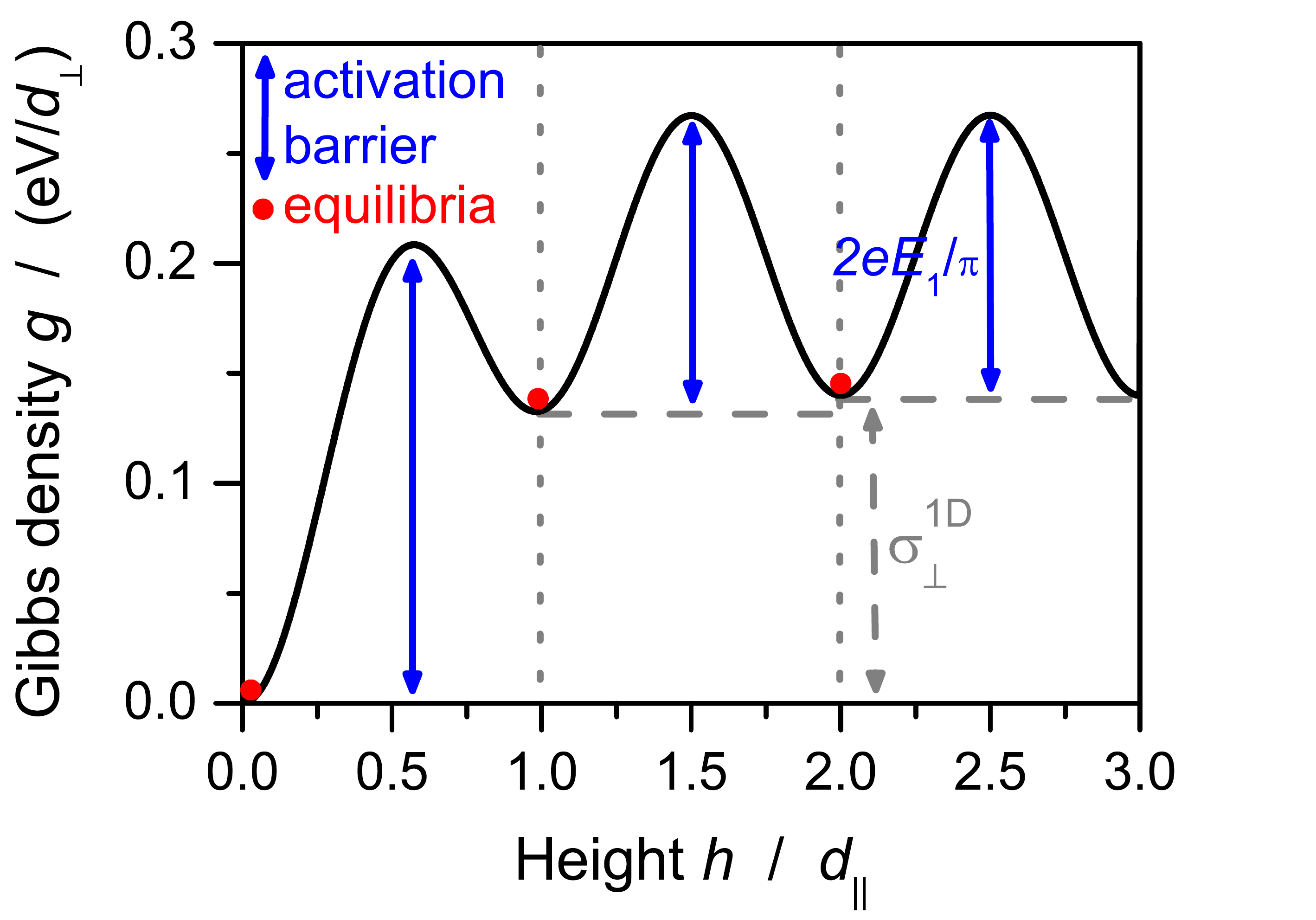}	
	\caption{ Homogeneous Gibbs free energy density $g_\text{hom}+2eE_0\tilde h/d_\perp$ (in units of $\text{eV}/d_\perp$) of a \ce{Li2O2} deposit versus dimensionless surface height $\tilde h=h/d_\|$ with molecule distance $d_\|$. The system is in metastable equilibrium at integer $\tilde h$. During growth of the first monolayer $h\le 1$, a nucleation barrier caused by the surface energy $\sigma_\perp$ must be overcome.}
	\label{fig:Gibbs}
\end{figure}

The thermodynamics of surface growth are defined by the \ce{Li2O2} free energy density, $g=g_b + g_s$ per substrate length. We estimate the bulk contribution as 
\begin{equation}
g_\text{b}=\frac{2e}{d_\perp\pi}\left[ -E_0 \pi \tilde{h} + E_1 \sin^2(\pi \tilde{h}) \right]
\end{equation}
where $\mu^\Theta = - 2e E_0$ is determined by the open circuit voltage. Our choice is motivated by the following: With complete molecular layers, {\it i.e.}, at integer ratios $\tilde{h}=h/d_\|$, the system is in equilibrium (Fig. \ref{fig:Gibbs}). The voltage barrier $E_1$ for homoepitaxial growth of a monolayer between these metastable equilibria accounts for the increased free energy of reaction intermediates (see Fig. \ref{fig:Exp}). The parameters $E_0$ and $E_1$ are taken from galvanostatic discharge measurements. We find the open circuit voltage $E_0 =2.96~\text{V}$ and the typical overpotential $E_1=0.2~\text{V}$, at which all reaction steps are downhill in energy \cite{Hummelshoj2010,Viswanathan2011}. We add Gaussian noise with standard deviation $0.004~\text{V} = 0.15\, k_\text{B}T/e$ to $E_1$ to model molecular fluctuations.  The microscopic surface energy density is $g^\text{micro}_\text{s}=\sigma_\perp^\text{1D}\left|\theta(h)\right|+\sigma_\|^\text{1D}\left|\frac{\partial h}{\partial x}\right|$.  Our continuous description,
\begin{equation}
g_s=\frac{1}{2} \left[ A \left( \sigma_\perp^\text{1D} + \sigma_\|^\text{1D} \right) +  \sigma_\perp^\text{1D} - \sigma_\|^\text{1D} \right ]   - \sigma_\perp^\text{1D}  e^{-\beta\tilde{h}^2/2}   
\end{equation}
smoothes the orientation-dependent surface energy (first term)~\cite{Stone2005}  and 
distributes the nucleation energy $\sigma_\perp^\text{1D}$ to initiate growth over a few monolayers with $\beta=6$ (second term).

The chemical potential then takes the dimensionless form
\begin{equation}
\tilde{\mu} = \tilde{\mu}_\text{hom}(\tilde{h}) - \tilde{\kappa}  \, 
\frac{\frac{\partial^2 \tilde h}{\partial \tilde x^2}}{  \left[1 +\left( \frac{\partial h}{\partial x} \right)^2\right]^{3/2}} 
\end{equation}
where $\tilde{\mu}=\mu/k_\text{B}T$ and $\tilde x = x/d_\perp$. 
The homogeneous term    
\begin{equation}
\tilde{\mu}_\text{hom} =  -\tilde{E}_0  + \tilde{E}_1 \sin(2\pi \tilde{h}) + \tilde{E}_2  \tilde{h}\, e^{-\beta\tilde{h}^2/2}
\end{equation}
describes a uniform film of $\tilde{h}=h/d_\|$ layers, where $E_2= \beta \sigma_\perp^\text{1D} d_\perp / 2e$ is the nucleation voltage to initiate heteroepitaxial growth and $\tilde{E}_i = 2eE_i/k_\text{B}T$. The inhomogeneous term reproduces the Cahn-Hilliard (CH) gradient expansion~\cite{cahn1958}, $\Delta\tilde{\mu} \sim -\tilde{\kappa} \frac{\partial^2 \tilde{h}}{\partial \tilde x^2}$, for small inclinations $\left| \frac{\partial h}{\partial x}\right| \ll 1$ with a dimensionless gradient energy penalty, $\tilde{\kappa} = ( \sigma_\perp^\text{1D} + \sigma_\|^\text{1D}) d_\|^2 / (d_\perp 2k_\text{B}T)$.  In contrast to the CH model, however, the gradient energy saturates at large inclinations.

The dynamics of surface growth follow from the theory of electrochemical nonequilibrium  thermodynamics~\cite{Bazant2013}, 
\begin{equation}
\frac{\partial \tilde c}{\partial \tilde t} - \frac{\partial}{\partial \tilde x}\left( \tilde M \tilde c \frac{\partial\tilde \mu}{\partial \tilde x}\right) = \tilde I(\tilde\mu,\Delta\tilde\Phi)
\end{equation}
where $\tilde{M}=M k_\text{B}T/(A_\perp d_\perp^2 k_0)$ is the dimensionless mobility for surface diffusion and $\tilde I=I/(2ek_0)$ is the dimensionless current density scaled to the exchange current density in the standard state ($a=1$). Since the dynamics is reaction limited, the dimensionless time, $\tilde{t} = t A_\perp k_0$, is scaled to the standard exchange time per surface site. This equation generalizes the CH and Allen-Cahn equations for electrochemistry. As in the case of anisotropic \ce{LiFePO4} nanoparticles~\cite{Singh2008}, diffusion can be neglected ($M=0$) to yield the Butler-Volmer Allen-Cahn reaction (ACR) equation~\cite{Bai2011,Bazant2013}, which, using Eqs. (\ref{eq:nernst})-(\ref{eq:A}) takes the dimensionless form,
\begin{equation}
\label{eq:growth rate}
\frac{\mathcal{D}\tilde{h}}{\mathcal{D}\tilde{t}} 
= \frac{\frac{\partial \tilde h}{\partial \tilde t}}{  \sqrt{ 1 +\left( \frac{\partial h}{\partial x} \right)^2}} 
= e^{-\alpha \Delta\tilde\Phi} - e^{(1-\alpha)\Delta\tilde\Phi+\tilde{E}_0+\tilde{\mu}}
\end{equation}
where $\Delta\tilde\Phi = 2e\Delta\Phi/k_\text{B}T$.  For galvanostatic discharge, the ACR equation is solved subject to the constraint of constant mean current density~\cite{Bai2011},  
\begin{equation}
\label{eq:mean rate}
\tilde{\bar{I}} = \frac{1}{L}\int_0^L \tilde{I} \text{d}x
\end{equation}
where $L$ is the substrate length.  Numerical integration of Eq. \ref{eq:growth rate} with periodic boundary conditions is performed in MATLAB employing the implicit DAE-solver \emph{ode15s}, and some analytical results are also possible.

\begin{figure}[t]
	\centering
	\includegraphics[width=\columnwidth]{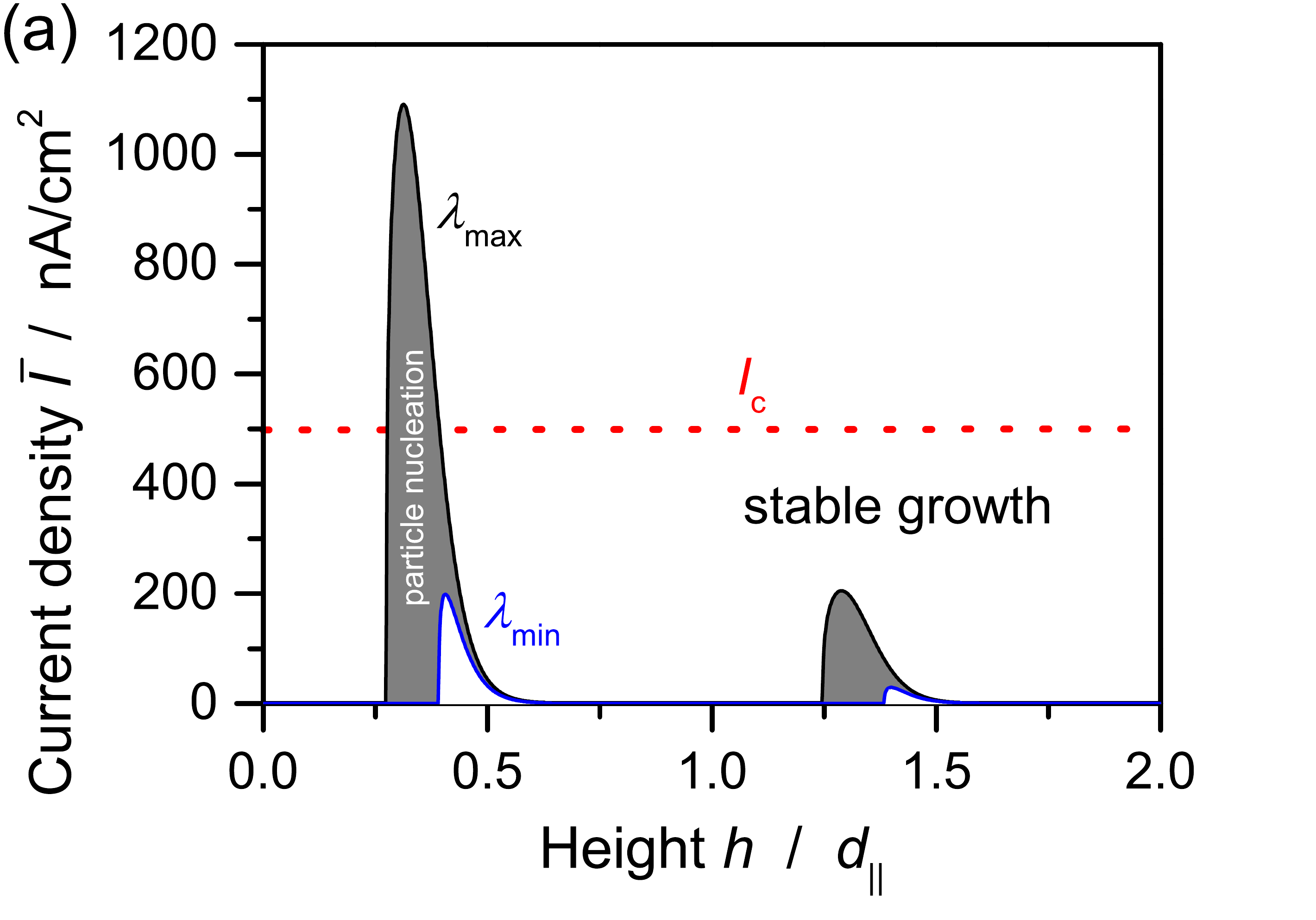}
	\includegraphics[width=\columnwidth]{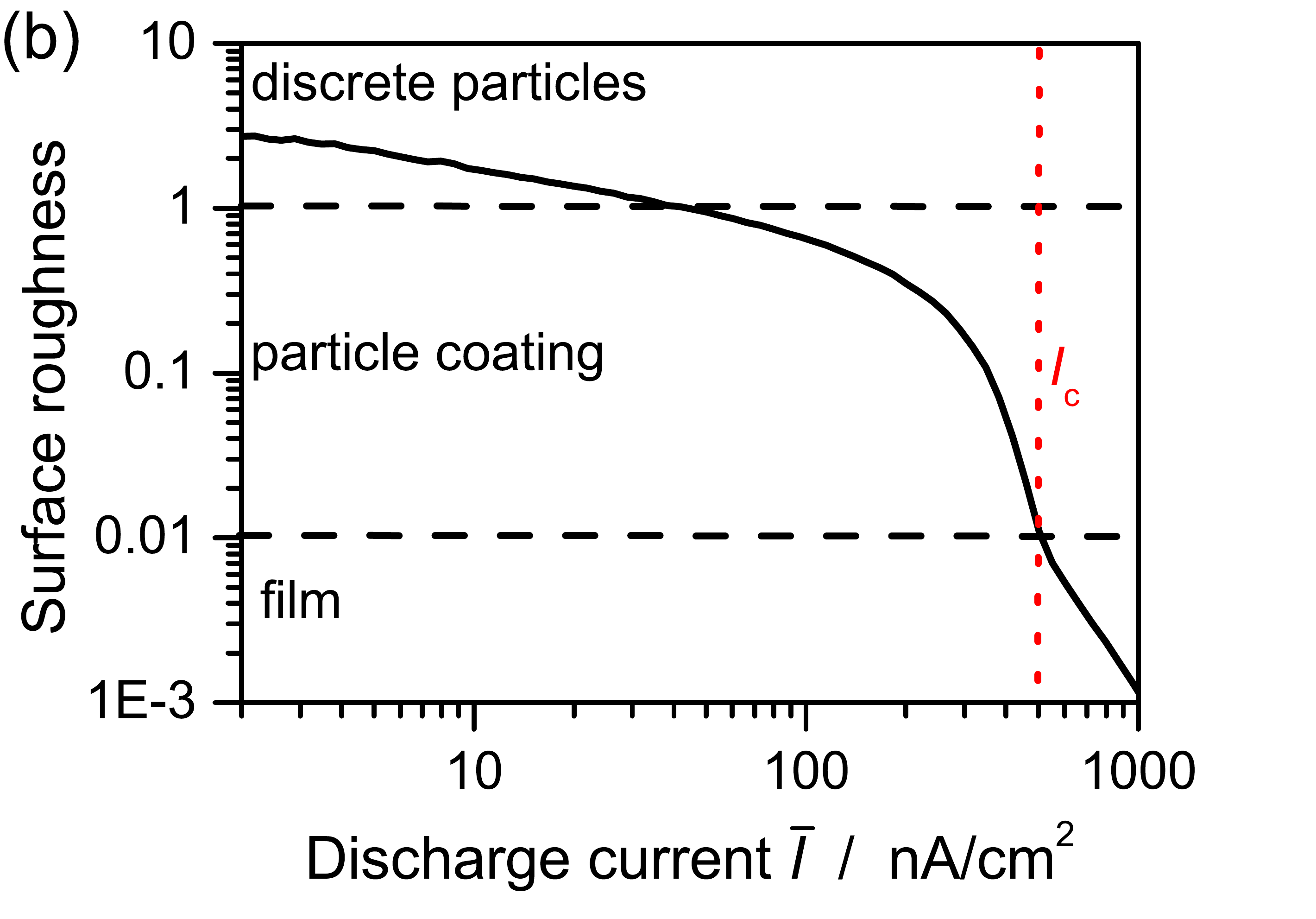}
	\caption{(a) Dependence of spinodal region on the applied current. The curves give the boundary between particle growth and film growth according to linear stability analysis, {\it i.e.}, $s(\tilde k;\tilde{\bar{I}})=\tilde{\bar{I}}$ (see Eq. \ref{eq:stability}). The black line corresponds to the most unstable wavelength $\lambda \rightarrow\infty$, the blue line to the most stable wavelength $\lambda=3d_\perp$. Dimensionless height $\tilde h=h/d_\|$ is shown. (b) Surface roughness after numerical evolution to mean height $\overline h=2 d_\|$. The standard deviation $\Delta[h]$ of $h(x)$ normalized by mean height $\overline h$ (see Eq. \ref{eq:surface roughness}) is depicted as a function of mean discharge rate $\bar I$ (see Eq. \ref{eq:mean rate}). The dashed lines illustrate the transition from growth of discrete particles over particle coating to film growth as a function of discharge current $\bar I$.}
	\label{fig:stability}
\end{figure}


{\bf Results.} The mechanism of rate-dependent morphology can be  understood by approximating Eq. \ref{eq:growth rate} in the linear and the Tafel regimes of small and large dimensionless overpotential, $\tilde{\eta} = 2e\eta/k_\text{B}T < 0$, respectively. Since we set $\gamma_\ddag=1$,  the \ce{Li2O2} chemical potential only influences the backward (dissolution) reaction.  At low rates, $\bar I\ll I_0$ or $|\tilde{\eta}|\ll1$, the forward (deposition) and backward reaction both contribute to the overall linear response, so the \ce{Li2O2} chemical potential drives the growth,
\begin{equation}
\label{linear regime}
\frac{\mathcal{D}\tilde{h}}{\mathcal{D}\tilde{t}} \sim - \tilde{\eta} = -( \Delta\tilde{\Phi}+\tilde{E}_0 + \tilde{\mu})
\end{equation}
Aside from the arc-length correction (left side), this is equivalent to the classical Allen-Cahn equation.  Analogous to spinodal decompositions, homogeneous growth becomes unstable when $\frac{\partial\tilde\mu}{\partial \tilde{h}}=0$, and particles develop. In the Tafel regime, far above the exchange current, $\bar I\gg I_0$ or $|\tilde{\eta}|\gg1$, the backward reaction is negligible, and the overall rate becomes independent of the chemical potential,
\begin{equation}
\frac{\mathcal{D}\tilde{h}}{\mathcal{D}\tilde{t}}  \sim e^{-\alpha \Delta\tilde\Phi}   \label{eq:tafel}
\end{equation}
enforcing film growth. In summary, the theory predicts a transition from particle to film growth with increasing discharge rate, analogous to the suppression of phase separation in \ce{LiFePO4} ~\cite{Bai2011,Cogswell2012}.

As with ion intercalation~\cite{Bai2011,Cogswell2012}, the transition in surface growth can be precisely identified by linear stability analysis. Fluctuations of dimensionless wave number $\tilde k=k d_\perp=2\pi/\tilde\lambda$ in a uniformly growing, homogeneous film, $\tilde{h}_0 = \tilde{\bar{I}} \tilde{t}$ (the base state), grow with the exponential rate,
\begin{eqnarray}
\label{eq:stability}
 \tilde s(\tilde k;\tilde{\bar{I}})&=&\frac{-\tilde{\bar{I}} }{\exp\left(-\tilde{\eta}_0\right)-1} \left[\frac{\partial\tilde\mu_\text{hom}}{\partial \tilde h}-\tilde k^2\frac{\partial\tilde\mu}{\partial \frac{\partial^2\tilde h}{\partial \tilde x^2}}\right] \\
&=& - \frac{ \tilde{\tilde{I}} \left[ \tilde E_1 2\pi\cos(2\pi \tilde h)
+\tilde E_2 (1-\beta\tilde h^2)  e^{-\beta \tilde h^2}+\tilde\kappa\tilde k^2 \right]}{\exp\left(-\tilde{\eta_0}\right)-1}  \nonumber
\end{eqnarray}
where $\tilde{\eta}_0$ is the overvoltage required for uniform growth, which solves $\tilde{I}(\tilde{h_0},\tilde{\eta}_0)=\tilde{\bar{I}}$. We derive this equation below (see Eq. \ref{eq:stability analysis result}). The dynamics are unstable ($\tilde s>0$) for all currents  if $\frac{\partial\mu}{\partial h}=\frac{\partial^2 g}{\partial h^2}<0$. Indeed, this occurs between the equilibria at full molecular layers (see Fig. \ref{fig:Gibbs}). Development of instabilities into particles requires that they grow faster than the homogeneous film, {\it i.e.}, $\tilde s>\tilde{\bar I}$. We evaluate this condition for marginal stability in Fig. \ref{fig:stability}a for the most unstable wavelength $\tilde\lambda\rightarrow\infty$ and the most stable wavelength $\tilde\lambda=3$ at which particles can still develop. Note that local noise favors small wavelengths. Above a critical current, growth will be homogeneous. This analysis overestimates the critical currents as it neglects the nonlinearity of the dynamics. The transition from particle growth to film growth is broad because of the strong dependence of the marginal stability on the wavelength of the fluctuation. Growth is most unstable during nucleation of the first monolayer when the nucleation energy $\tilde E_2$ must be overcome. Thus, at intermediate currents, nucleation of particles can be followed by homogeneous growth at thicker coatings. 

The numerical stability analysis shown in Fig. \ref{fig:stability}b confirms this picture. Far below the exchange current, the growth of distinct particles is signaled by normalized standard deviations of the height profile $h(x)$ larger than unity.  Above the exchange current, a tiny surface roughness signals film growth. An intermediate regime of particle coatings separates these extremes.

\begin{figure}[t]
	\centering
	\includegraphics[width=\columnwidth]{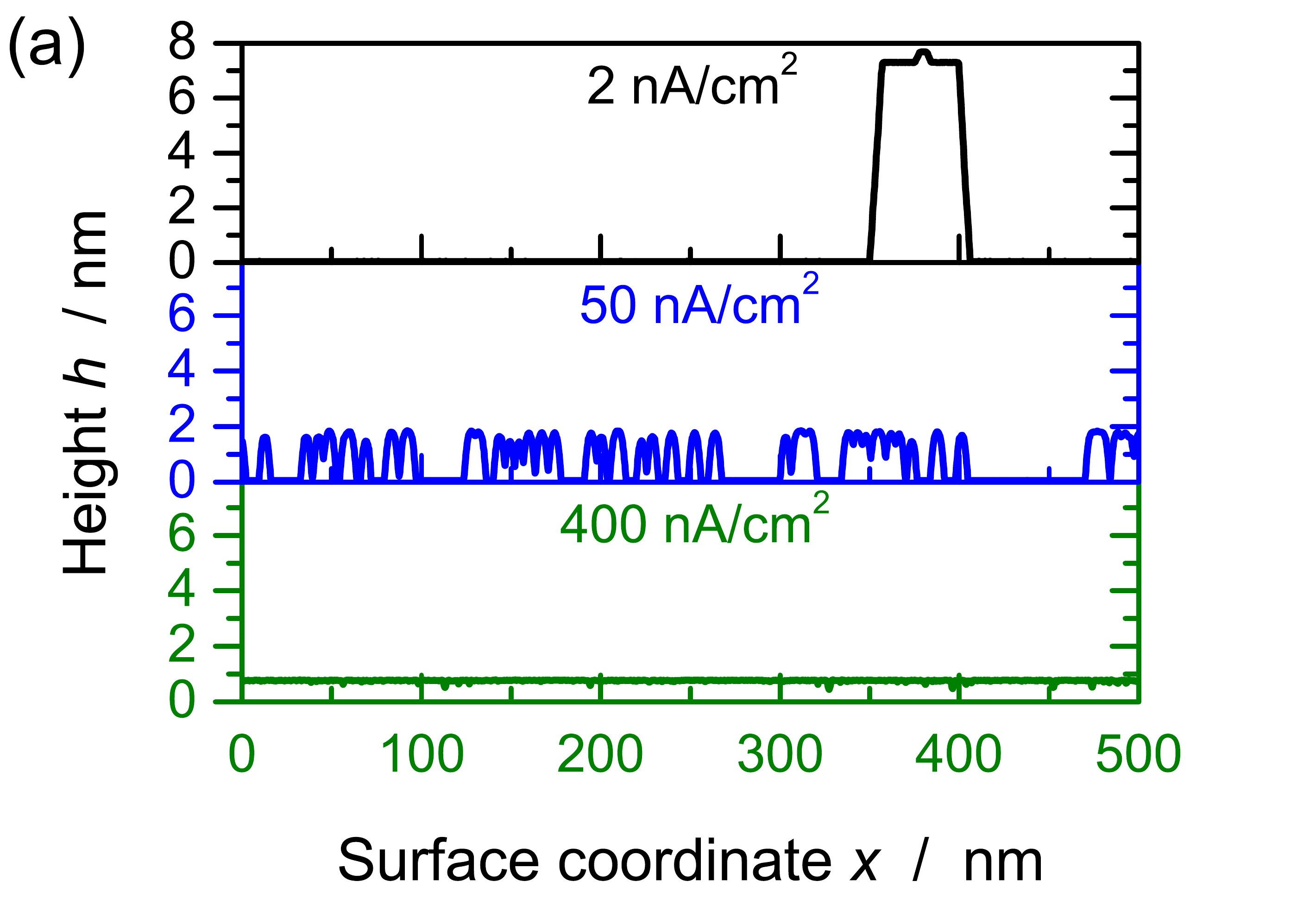}
	\includegraphics[width=\columnwidth]{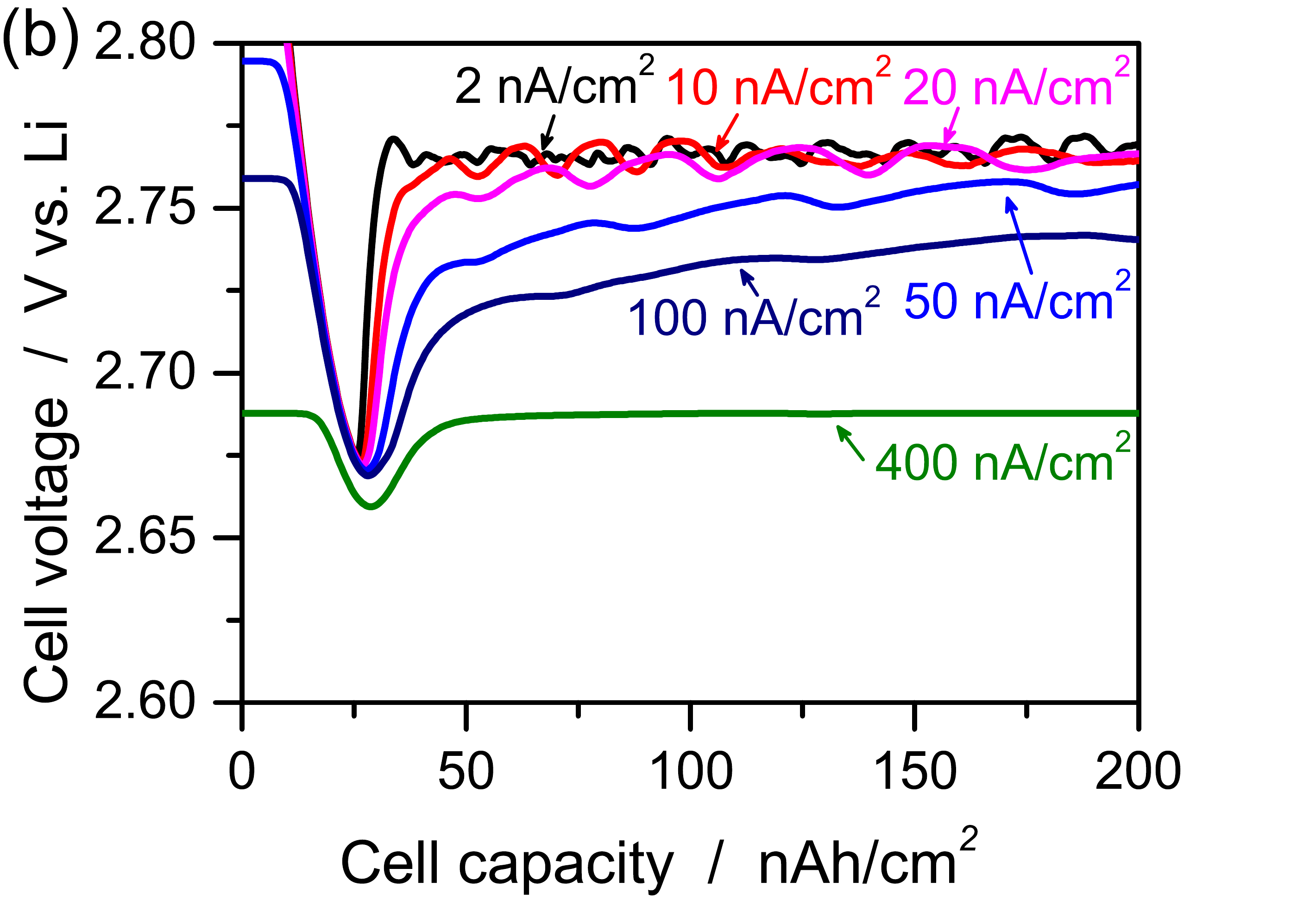}
	\caption{Simulated surface growth for various discharge currents $\bar{I}$. (a) Height profile in during galvanostatic discharge to two molecular monolayers. The growth mode undergoes a transition from particle to film growth with increasing rates. (b) Cell potential during galvanostatic discharge. The dip corresponds to the nucleation process.}
	\label{fig:model}
\end{figure}

\begin{figure}[t]
	\centering
	\includegraphics[width=0.8\columnwidth]{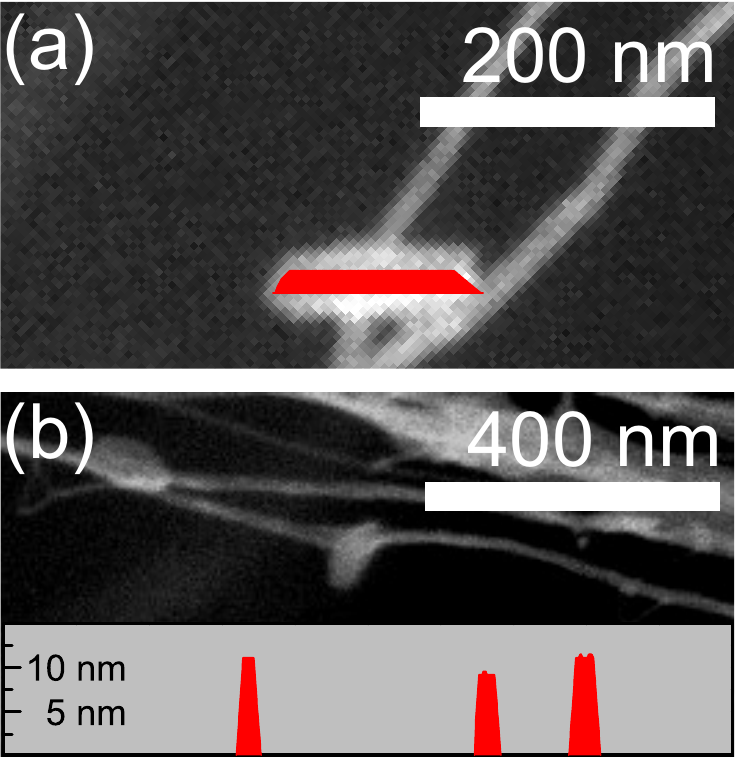}
	\caption{(a) Validation of disc-like particle morphologies realizing the Wulff-shape at $I=2 \text{ nA/cm}^\text{2}$ \cite{Mitchell2013a}. The shaded area shows a modeled disc shape. The aspect ratio in TEM micrograph and model agrees. (b) The average particle distance is in the same order of magnitude, $500 \text{ nm}$, in SEM micrograph and theory.}
	\label{fig:validation}
\end{figure}

The exchange current density $I_0$ is determined via Tafel analysis \cite{Gallant2013}. We must carefully interpret this measurement because the exchange current density depends on \ce{Li2O2} activity and height profile in our model (see Eq. \ref{eq:I0}). Experimental Tafel analysis adjusts the Tafel slope to match the kinetics of uniform growth at large rates, which is described by Eq. \eqref{eq:tafel}. The current is then extrapolated from the large overpotential regime to zero overpotential $\Delta\Phi=0$, yielding the value $I_0^\Theta=2ek_0=2 \text{ nA/cm}^\text{2}$. It corresponds to the exchange current density in the thermodynamic standard state, where $a=1$. The thermodynamic standard state is the fully charged battery without \ce{Li2O2}, {\it i.e.}, $h(x)=0$. The instabilities, however, develop close to the spinodal point, $h\approx d_\|/4$. Therefore, the critical current for the transition in \ce{Li2O2} morphology is the exchange current density evaluated at the spinodal point, {\it i.e.}, 
\begin{equation}
I_c=I_0(h = d_\|/4)=I_0^\Theta a_{d_\|/4}^\alpha=500 \text{ nA/cm}^\text{2}.
\end{equation} 
This exchange current agrees with the transition current predicted by linear stability analysis as demonstrated in Fig. \ref{fig:stability}. Our Tafel analysis gives the symmetry factor $\alpha=0.1$ on CNTs \cite{Gallant2013} and $\alpha=0.2$ on glassy carbon \cite{Lu2013}. The small apparent symmetry factor observed on CNTs could stem from additional overpotentials, e.g., diffuse double layers in the solid due to low electron conductivity in \ce{Li2O2} \cite{Bazant2005, Biesheuvel2009}. Therefore, we evaluate our model for the theoretical value $\alpha=0.25$ as discussed above. 

These parameters allow the quantitative comparison between model and experiment. Electron mircoscopy images of \ce{Li2O2} on CNT electrodes during galvanostatic discharge are shown in Fig. \ref{fig:Exp}b/c \cite{Mitchell2011, Gallant2012a, Mitchell2013a, Gallant2013}. The predictions of our surface growth model are summarized in Fig. \ref{fig:model}a. At very low surface specific discharge rates $2 \text{ nA/cm}^\text{2}\ll I_0$, distinct disc-like particles nucleate and evolve into toroid-like ones (compare with Fig. \ref{fig:Exp}b). At intermediate rates $50 \text{ nA/cm}^\text{2} < I_0$, small particles are coating the CNTs (compare with Fig. \ref{fig:Exp}c). At very large rates $400 \text{ nA/cm}^\text{2}\gtrsim I_0$, a film is coating the CNTs. This prediction is in excellent agreement with the films observed at $1000\text{ nm/cm}^2$ on glassy carbon in Ref. \cite{Viswanathan2011}. 

Next, we validate cell voltages as shown in Fig. \ref{fig:Exp}a and Figs. \ref{fig:model}. Note that the simulations start at nonzero currents and overvoltages. The cell voltage goes through a minimum when $1/4$ molecular monolayers are formed and the system becomes unstable $\frac{\partial\mu}{\partial h}<0$ (see Eq. \ref{eq:stability}). The dip in cell potential is determined by the nucleation energy at low rates, {\it i.e.}, $\sigma_\perp^\text{1D}$. It is a bit smaller in experiment than in theory, possibly due to surface defects, averaging over numerous CNTs and surface capacities. Due to our choice of the symmetry factor, $\alpha=0.25$, overvoltages are generally too low, which may also reflect neglected transport and reaction processes in the solid. 

Finally, we analyze the predicted particle shape and particle density at very low currents. Our theory explains the presence of disc-shaped particles at low rates and capacities. These were found to be precursors of aggregated toroid-like particles \cite{Mitchell2013a}. The aspect ratios $\sigma_\perp^\text{1D}/\sigma_\|^\text{1D}=0.15$ found in our simulation (see Fig. \ref{fig:model}) agree with the theoretical Wulff shape and the values observed by TEM microscopy in Ref. \cite{Mitchell2013a}. We demonstrate this by continuing our simulations to larger capacities, at which individual discs can be imaged (see Fig. \ref{fig:validation}a). Furthermore, the predicted average particle distance of roughly $500 \text{ nm}$ is consistent with experimental imaging. 

{\bf Conclusion.} In this Letter we have developed a theory of electrodeposition based on non-equilibrium thermodynamics, combining existing models for surface growth and electrochemical reaction rates. The model quantitatively describes the transition from film growth to particle growth of \ce{Li2O2} during galvanostatic discharge of an \ce{Li}-\ce{O2} battery with increasing current. The predicted transition takes place around the exchange current $I_c$ of the oxygen reduction reaction at the nucleation barrier for growth of the first monolayer, which is two orders of magnitude larger than the exchange current $I_0^\Theta$ from Tafel analysis of high-rate film growth, as observed in experiments. 

Our theoretical framework for electrochemically-driven surface growth could be applied to other systems, such as \ce{Na}-\ce{O2} batteries \cite{Hartmann2012}, or extended to further dynamical regimes. After the initial phase of \ce{Li2O2} particle nucleation analyzed here, the particle morphology evolves from disc-like to toroid-like under certain conditions \cite{Granasy2005}, that may be describable by our approach, e.g. by including electron transport and elastic strain. Understanding these principles of \ce{Li2O2} crystallization is important for overcoming cell performance limitations due to the low electronic conductivity of \ce{Li2O2}.

The morphological transition from heterogeneous to homogeneous at a critical rate is a general prediction of the variational theory of chemical kinetics~\cite{Bazant2013}.  Using the same theory for reaction-limited dynamics of a concentration variable, $\tilde{c}=\tilde{h}$, such a transition was first predicted for lithium intercalation in \ce{Li_XFePO4} nanoparticles~\cite{Bai2011}, as the suppression of phase separation into \ce{LiFePO4} and \ce{FePO4} domains. The only difference lies in the thermodynamics of intercalation, given by a Cahn-Hilliard regular solution model~\cite{Singh2008}.  \ce{Li_XFePO4} intercalation is predicted to be stable and uniform above a critical current $I_c(X)$, somewhat below the typical Tafel exchange current due to coherency strain~\cite{Cogswell2012}.  In contrast, \ce{Li2O2} growth is always unstable, but transitions from high to low surface roughness at a critical current far above the Tafel exchange current.  In both cases, however, the transition occurs close to the exchange current at the spinodal point due to the exponential (Arrhenius, Butler-Volmer) dependence of the reaction rate on the local overpotential, or free energy of reaction.

\vspace{0.1in}

{\small 
{\bf Methods.}  In this section, we provide additional mathematical details.  First, we derive the exponential growth rate for linear instability in Eq. \ref{eq:stability}. We decompose total surface height $\tilde h=\tilde h_0 +\delta \tilde h_{\tilde{k}}$ into height of the uniform film $h_0$ and of fluctuations $\delta \tilde h_{\tilde{k}}$ of wavenumber ${\tilde k}$. Their second derivative is $\frac{\partial^2 \delta\tilde h }{ \partial \tilde x^2}=-\tilde{k}^2\delta \tilde h$. $\delta A=0$ vanishes because $A$ depends on $h$ through the square of $\frac{\partial h }{ \partial x}$ only. In order to determine $\delta (\Delta\Phi)_{\tilde{k}}$, we study the effect of fluctuations in surface height on the mean discharge current in Eq. \ref{eq:mean rate}
\begin{eqnarray}
\label{eq:fluctuation mean current}
0&=&\delta \tilde{\bar I}=\frac{1}{L}\int_0^L \delta \tilde I \text{d}x \nonumber \\
&=&-\delta(\Delta\tilde\Phi)_{\tilde{k}}\left[\alpha e^{-\alpha \Delta\tilde\Phi_0} + (1-\alpha) a(\tilde h_0) e^{(1-\alpha)\Delta\tilde\Phi_0}\right] \nonumber\\
&&-  \frac{(1-\alpha) a(\tilde h_0) e^{(1-\alpha)\Delta\tilde\Phi_0}}{L}\int_0^L\delta\tilde \mu_{\tilde{k}}\text{d}x,
\end{eqnarray}
where $\Delta\tilde\Phi_0$ is the voltage step required for uniform growth, which solves $\tilde{I}(\tilde{h_0},\tilde{\eta}_0)=\tilde{\bar{I}}$. The integral
\begin{equation}
\int_0^L\delta\tilde \mu\text{d}x= \left[\frac{\partial\tilde\mu}{\partial \tilde h}-\tilde k^2\frac{\partial\tilde\mu}{\partial \frac{\partial^2\tilde h}{\partial \tilde x^2}}\right]\int_0^L\delta\tilde h_{\tilde{k}}\text{d}x=0
\end{equation}
vanishes for all $\tilde k>0$. Therefore, according to Eq. \ref{eq:fluctuation mean current}, $\delta(\Delta\tilde\Phi)_{\tilde k}=0$ vanishes, too. We can now calculate the dynamics of the fluctuations $\delta\tilde h_{\tilde{k}}$ from Eq. \ref{eq:growth rate}
\begin{equation}
\label{eq:tmp stability analysis}
 \frac{\partial \delta\tilde h_{\tilde{k}}}{\partial \tilde{t}} =- \delta \tilde{h}_{\tilde{k}} a(\tilde h_0) e^{(1-\alpha)\Delta\tilde\Phi_0} \left[\frac{\partial\tilde\mu_\text{hom}}{\partial \tilde h}-\tilde k^2\frac{\partial\tilde\mu}{\partial \frac{\partial^2\tilde h}{\partial \tilde x^2}}\right].
\end{equation}
We want to substitute $\tilde{\bar I}$ and $\tilde{\eta}_0$ for $a(\tilde h_0)$ and $\Delta\tilde\Phi_0$. To this aim, we write for the homogeneous base state
\begin{eqnarray}
\tilde{\bar I} &=& e^{-\alpha \Delta\tilde\Phi_0} - a(\tilde h_0) e^{(1-\alpha)\Delta\tilde\Phi_0} \nonumber\\
&=&a(\tilde h_0)e^{(1-\alpha)\Delta\tilde\Phi_0} \left[e^{-\Delta\tilde\Phi_0-\tilde{E}_0-\tilde{\mu}(\tilde h_0)}-1\right] \nonumber\\
\tilde{\bar I}&=&a(\tilde h_0)e^{(1-\alpha)\Delta\tilde\Phi_0} \left[e^{-\tilde\eta_0}-1\right]
\end{eqnarray}
and rewrite Eq. \ref{eq:tmp stability analysis}
\begin{equation}
 \frac{\partial \delta\tilde h_{\tilde{k}}}{\partial \tilde{t}} =   \frac{-\tilde{\bar{I}} \delta \tilde{h}_{\tilde{k}}}{\exp\left(-\tilde{\eta}_0\right)-1} \left[\frac{\partial\tilde\mu_\text{hom}}{\partial \tilde h}-\tilde k^2\frac{\partial\tilde\mu}{\partial \frac{\partial^2\tilde h}{\partial \tilde x^2}}\right].
\end{equation}
The exponential growth rate in Eq. \ref{eq:stability} is 
\begin{equation}
\label{eq:stability analysis result}
\tilde s(\tilde k;\tilde{\bar I})=\frac{\frac{\partial \delta\tilde h_{\tilde{k}}}{\partial \tilde{t}}}{\delta\tilde h_{\tilde{k}}}.
\end{equation}
The marginal stability curve in Fig. \ref{fig:stability} is determined by solving $\tilde s=\tilde{\bar I}$ for $\exp\left(-\tilde{\eta}_0\right)$ and substituting into Eq. \ref{eq:current}.

In Fig. \ref{fig:stability}, we determine surface roughness $\Delta [h]$ as normalized standard deviation of $h(x)$ according to
\begin{equation}
\label{eq:surface roughness}
\Delta [h] = \sqrt{\frac{1}{L}\int_0^L \frac{\left(h(x)-\bar{h}\right)^2}{\bar{h}^2}\text{d}x}
\end{equation}
with the mean height
\begin{equation}
\bar h = \frac{1}{L}\int_0^L h(x)\text{d}x.
\end{equation}

We numerically integrate the DAE system of Eq. \ref{eq:growth rate} and Eq. \ref{eq:mean rate} in MATLAB employing the DAE-solver \emph{ode15s}. It is an implicit, variable order solver. Periodic boundary conditions are used. Spatial derivatives are calculated with first order central differencing. The spacing of grid points is given by the distance between molecules $d_\perp$. Simulations were performed in systems of length $L=500\text{ nm}$ (Figs. \ref{fig:stability},\ref{fig:model},\ref{fig:validation}a) and $L=1000\text{ nm}$ (Fig. \ref{fig:validation}b).

{\bf Acknowledgments.} This work was suppported in part by MRSEC Program of National Science Foundation under award number DMR-0819762. B.H. acknowledges support from the German Academic Exchange Service (DAAD). B.M.G. acknowledges a National Science Foundation Graduate Research Fellowship. The authors would like to acknowledge Carl V. Thompson for fruitful discussions.

}


\bibliography{library}

\end{document}